\documentclass[journal=jacsat,manuscript=traditional,layout=onecolumn]{achemso}
\usepackage{titlesec}

\titleformat{\section}{\normalfont\Large\bfseries}{\thesection}{1em}{}
\titleformat{\subsection}{\normalfont\normalsize\bfseries}{\thesubsection}{1em}{}
\titleformat{\subsubsection}{\normalfont\normalsize\itshape}{\thesubsubsection}{1em}{}

\usepackage[version=3]{mhchem} 
\usepackage{xcolor}
\usepackage[pagewise]{lineno}
\usepackage{xr}
\usepackage{hyperref}\hypersetup{
  colorlinks   = true, 
  urlcolor     = blue, 
  linkcolor    = blue, 
  citecolor   = blue 
}
\usepackage{graphicx}
\usepackage{comment}

\makeatletter

\newcommand*{\addFileDependency}[1]{
\typeout{(#1)}
\@addtofilelist{#1}
\IfFileExists{#1}{}{\typeout{No file #1.}}
}\makeatother

\newcommand*{\myexternaldocument}[1]{%
\externaldocument{#1}%
\addFileDependency{#1.tex}%
\addFileDependency{#1.aux}%
}
\myexternaldocument{00_SI}

\author{Suman Bhattacharjee}
\affiliation{Centre for Research in Nanotechnology \& Science (CRNTS), Indian Institute of Technology Bombay, Mumbai-400 076, India}
\author{Sanjoy Khawas}
\affiliation{Soft Matter and Nanomaterials Laboratory, Department of Physics, Indian Institute of Technology Bombay, Mumbai-400 076, India}
\author{Sunita Srivastava}
\affiliation{Soft Matter and Nanomaterials Laboratory, Department of Physics, Indian Institute of Technology Bombay, Mumbai-400 076, India}
\email{sunita.srivastava@iitb.ac.in}

\title{Tunable Nanoparticle Stripe Patterns at Inclined Surfaces\footnote{Preprint}}
 
\begin{document}
\newpage
\begin{abstract}

Periodic assemblies of nanoparticles are central to surface patterning, with applications in biosensing, energy conversion, and nanofabrication. Evaporation of colloidal droplets on substrates provides a simple yet effective route to achieve such assemblies. This work reports the first experimental demonstration of patterns formed through stick–slip dynamics of the three-phase contact line during evaporation of gold nanoparticle suspensions on inclined substrates. Variation in nanoparticle concentration and substrate inclination alter the balance of interfacial and gravitational forces, producing multiple stick–slip events that generate periodic stripes. Stripe density exhibits a sinusoidal dependence on inclination angle, while inter-stripe spacing remains nearly invariant. Independent control over inter-stripe spacing is achieved through adjustment of nanoparticle or surfactant concentration. These results highlight the complex interplay of gravitational and interfacial forces in directing periodic nanoparticle assembly and establish a versatile, programmable framework for surface patterning with tunable nano/microscale dimensions.
\end{abstract}

\section{Introduction}
The ability to direct nanoparticles into well-defined patterns through self-assembly plays a vital role in developing next-generation nanotechnologies. Droplet evaporation-based self-assembly offers a simple, low-cost approach for the creation of ordered structures \cite{guerrero2009gemini,srivastava2020dual,feng2022long,zargartalebi2022self,elbert2022evaporation,khawas2024directing}, though challenges in reproducibility and structural control persist. 
Ordered ring or stripe patterns of colloidal particles \cite{watanabe2009mechanism,srivastava2020dual} are central to applications in biosensing \cite{li2011colloidal}, energy conversion \cite{zhang2013nanomaterials}, and advanced printing and patterning technologies \cite{de2004inkjet,winhard2021direct}, and are the focus of this study. The formation of stripe patterns is intricately linked to the dynamics of the three-phase contact line (TPCL) during solvent evaporation. The TPCL undergoes a series of pinning-depinning events, which significantly influence the final deposition pattern, often leading to the formation of multiple rings or stripes besides a primary coffee-ring.\par
A diverse array of experimental methodologies in sessile droplet evaporation has been utilized to fabricate stripe patterns through the modulation of the stick-slip dynamics of the TPCL \cite {yang2014multi,seo2017altering}. Kim and collaborators used high-speed interferometry to directly observe the emergence of stripe patterns as a result of stick-slip activities under the influence of particle wettability\cite{kim2018effect}. Srivastava \textit{et al.} successfully generated concentric ring structures on hydrophilic silicon (\textit{Si}) surfaces by varying the concentration levels of DNA-coated gold nanoparticles\cite{srivastava2020dual}. Formation of multi-ring or stripe patterns has been reported using other methods like the capillary bridge technique\cite{xu2006self}, convective self-assembly\cite{watanabe2009mechanism,mino2015situ}, ``Ball-on-film" method\cite{zhang2016evaporation}. 
\par
Despite prior reports of stripe-like deposits, they offer limited control over stripe dimensions, and the role of gravity in enabling such control remains unexplored. The gravitational influence on inclined droplets and the TPCL dynamics have been a subject of sustained interest, with early theoretical and experimental studies exploring how gravity deforms droplet shape and alters contact line behavior \cite{brown1980static,rotenberg1984shape,roura2001equilibrium,espin2014sagging,xie2016droplet,mamalis2016motion,lv2018wetting,al2018water,ravazzoli2019contact,timm2019evaporation,cai2023asymmetric}. Early works from Qu{\'e}r{\'e} \textit{et al.} demonstrated that droplets can remain pinned on a tilted surface as long as contact angle hysteresis exists, and showed that in the case of small hysteresis, droplet shapes remain close to spherical caps despite the incline \cite{quere1998drops}. These observations were further supported by studies that established linear correlations between the droplet aspect ratio and the extent of contact angle hysteresis \cite{dhar2020surface}, helping to quantify the degree of gravitational flattening in relation to surface wettability. A number of computational models have investigated how gravity influences evaporation behavior and solute redistribution. Du and Deegan simulated TPCL deposit variations under tilt using a two-dimensional strip-like droplet approximation \cite{du2015ring}. A very recent study by D'Ambrosio \textit{et al.}\cite{d2025effect} provided theoretical predictions for the lifetime of sessile and pendant droplets evaporating in various modes of evaporation, such as constant radius, constant contact angle, stick-slip, and stick-jump.
\par
Although there is ample theoretical literature on the gravitational effect on an inclined pure water droplet\cite{chou2012drops,de2021shape,pan2022evaporation,chen2023simulation}, studies on particle-laden droplets under gravity require attention so as to understand the dried pattern formations. Sessile droplets on inclined substrates exhibit distorted shapes and uneven evaporation rates compared to their horizontal counterparts, often undergoing more frequent transitions between evaporation modes \cite{varagnolo2013stick,kim2017evaporation}. This results in inhomogeneous deposition features, most notably the formation of asymmetric coffee-ring structures, ``coffee-eye" or a non-homogeneous deposition\cite{gopu2020evaporation,kim2021evaporation,parsa2023inclined,mondal2018patterns,bansal2018beyond}. Varagnolo \textit{et al.} reported stick-slip motion of water droplets on a chemically heterogeneous surface\cite{varagnolo2013stick}, but this study lacks any surface patterning possibly due to  absence of solutes. In a recent study, Winahard \textit{et al.} demonstrated uniform deposition rising from a homogeneous inclined substrate, as a nanolitre-sized colloidal drop undergoes stick-slip mode of evaporation.\cite{winhard2021direct} Stripe patterns were not observed in this case, as the slip time was much greater compared to the stick phase. Beyond fundamental studies, gravitationally influenced evaporation has also found relevance in applied contexts. Hampton \textit{et al.} demonstrated directed colloidal structure formation on tilted surfaces \cite{hampton2012influence}, while studies on blood droplet drying showed that evaporation geometry affects deposition features relevant to lipid detection and forensic analysis \cite{hidalgo2024dried}. However, controlled patterning using inclined substrates remains largely unreported.
\par
In this study, we introduce a gravity-assisted self-assembly method that produces semi-circular strips of spherical gold nanoparticles (AuNPs) with controllable dimensions on silicon substrates. Precise control over nanoparticle concentration and the droplet’s evaporation dynamics on a hydrophilic \textit{Si} substrate is shown to be critical for inducing the oscillatory motion of the TPCL, which leads to the formation of multiple semi-circular stripe patterns within the primary coffee-ring. The oscillatory motion of the TPCL arises from the interplay between surface tension and gravitational forces. The number of stripes formed per unit length ($\sigma$) and inter-stripe spacings ($\lambda$) can be systematically tuned by varying the substrate tilt angle ($\phi$) and adjusting the concentrations of nanoparticles ($C_{np}$). This work establishes a clear correlation between gravitational and interfacial forces in controlling the resulting stripe morphology. Although previous studies have shown the effect of gravity on inclined colloidal drops, the controlled manipulation of pattern dimensions and the underlying mechanism have not been addressed, which is the primary focus of this paper. Furthermore, to the best of our knowledge, the use of gravity and surface tension to create controllable patterns has not yet been reported.

\section{Results and Discussion}
\subsection{Droplet evaporation on an inclined plane}

\begin{figure*}[h]
\centering
\includegraphics[width = {\textwidth}]{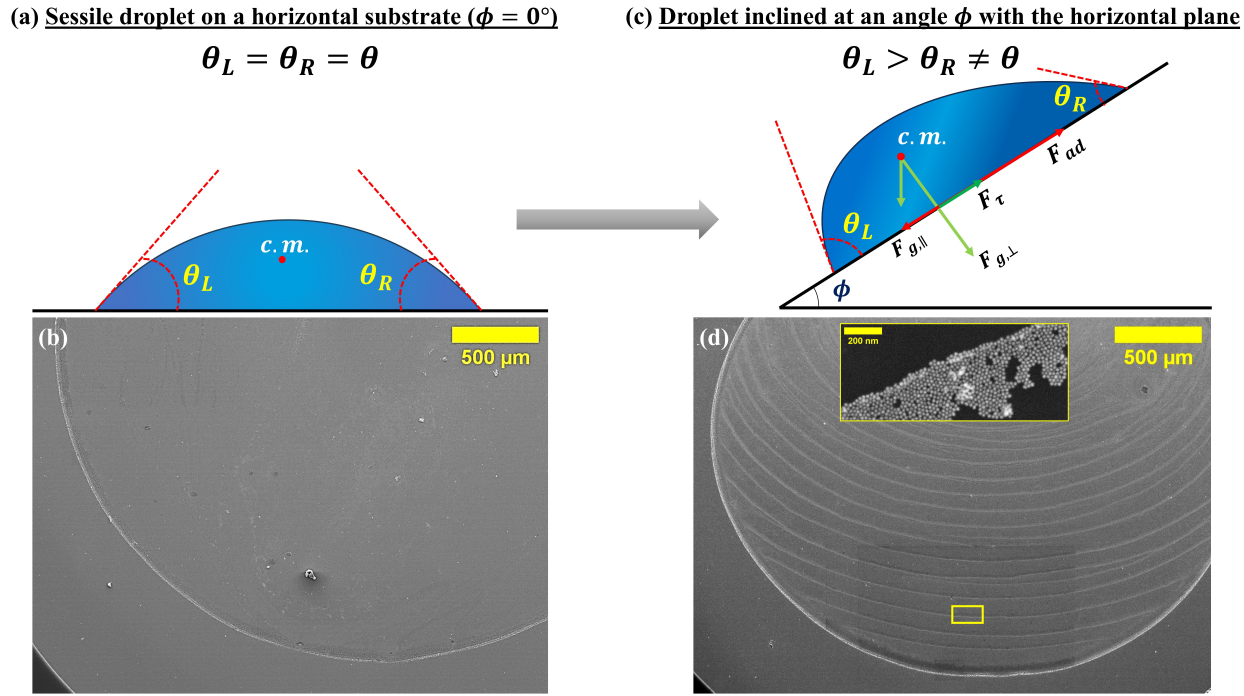}
\caption{Transition from (a) sessile to (c) inclined droplet configuration resulting in a shift in the center of mass (c.m.). Forces acting on the TPCL, along with the particle flow velocities, are shown. Dried deposit resulting in a coffee-ring (b) for a sessile drop, and ordered stripe patterns after evaporation (d). A typical representation of AuNP deposition on a single stripe (rectangular box) is shown via a high-magnification SEM image in the inset of (d).}
\label{tilt-schematic}
\end{figure*}

In Figure \ref{tilt-schematic}, the schematic depicting the experimental setup for droplet evaporation of the colloidal suspension and the corresponding SEM micrographs are shown. In a typical experiment, a $2~\mu L$ colloidal suspension is evaporated on a substrate inclined at various angles as shown in Figure \ref{tilt-schematic}. The details for nanoparticle synthesis protocol \cite{kumar2024shape} and experimental methodology are mentioned in the \nameref{sec:expt}. A sessile droplet of colloidal AuNPs on a flat hydrophilic substrate takes the shape of a spherical cap with equal contact angles on both sides of the pinned TPCL $(\theta~=\theta_L=\theta_R)$, as shown in Figure \ref{tilt-schematic}(a). Here, we define the pinned site contact (left) angle as $\theta_L$ and the rear side (receding side) contact angle as $\theta_R$.
The macroscopic sessile drop geometry in equilibrium is governed by Young's equation\cite{tatyanenko2025line}, 
\begin{equation}\label{eqn:Young}
    \gamma_{LV} cos\theta = \gamma_{SV} - \gamma_{SL} 
\end{equation}
where $\gamma_{AB}$ denotes the interfacial tensions between the liquid-vapour (\textit{LV}), solid-vapour (\textit{SV}), and solid-liquid (\textit{SL}) interfaces, respectively. When an inclination angle $\phi$ is introduced relative to the horizontal plane, the droplet tilts and deforms under gravity, leading to a shift in the center of mass (c.m.) towards the direction of the tilt, and results in an asymmetric droplet shape as shown in Figure \ref{tilt-schematic}(b). The shape deformation of the droplet can be correlated with $\Delta \theta = \theta_L - \theta_R$. In case of $\phi = 0^\circ$, $\Delta \theta \approx 0^\circ$, whereas as shown in Figure \ref{tilt-schematic}(b), for $\phi = 60^\circ$, the estimate of $\Delta \theta$ in experiments was found to be $6^\circ$. 
A standard coffee ring pattern for a droplet on a substrate with $\phi=0^\circ$ is depicted in the SEM micrograph of the dried deposit shown in Figure \ref{tilt-schematic}(b). In contrast, the dried pattern for a droplet on a tilted substrate, illustrated in Figure \ref{tilt-schematic}(d), displays periodic curved stripes. Hence, here we demonstrate an easy pathway for the creation of tunable microscale patterns with colloidal nanoparticles. Even though particle deposition on inclined substrates has been studied earlier\cite{winhard2021direct,kim2021evaporation}, the formation of such controlled period patterns using nanocolloids has not been reported. Qualitatively, the stripe pattern arises from the stick-slip motion of the TPCL at the elevated edge of the droplet. As the colloidal droplet is tilted, the contact angles gradually decrease during evaporation, triggering geometrical instability. This instability enables the receding section of the TPCL to undergo stick-slip motion \cite{srivastava2020dual}, allowing the formation of AuNP monolayer stripes on the \textit{Si} substrate [Figure \ref{tilt-schematic}(d)]. Nevertheless, a detailed quantitative understanding of the driving forces behind this instability is necessary to utilize this approach for programmable and tunable periodic patterning,  which constitutes the focus of the remainder of this paper.\par
We successfully fabricated microscale AuNP stripes using the inclined droplet evaporation technique; however, questions remain about the controlled tunability of the dimensional features of the resulting patterns. For a sessile drop, gravity exerts a downward force equal to the drop's weight ($mg$) acting vertically through its center of mass (c.m.). When the substrate is inclined, there is an effective component of gravity proportional to $\sin\phi$ [Figure \ref{tilt-schematic}(c)]. Such a force component has been reported to result in sliding/slipping of the droplet based on solution-substrate interaction\cite{al2020adhesion,winhard2021direct}. Interestingly, in our experiments, the TPCL on one side stayed pinned throughout most of the evaporation process, maintaining long-term droplet stability without sliding or slipping. Although the gravitational drag acts through the shifted c.m. of the bulk droplet, the translated component of this force along the TPCL ($F_{g,\parallel}$) governs the stick–slip motion. The force balance for a static inclined droplet through the shifted c.m. can be expressed as \cite{al2020adhesion},
\begin{equation}
\label{eqn:force balance}
   F_{g,\parallel}-F_\tau -F_{ad}=0
\end{equation}     
where $F_{g,\parallel}=mg\sin\phi$ is the gravitational force component along the TPCL, $F_\tau$ is the shear force at the solid–liquid interface, and $F_{ad}$ is the adhesion force due to interfacial interactions. While $F_{g,\parallel}$ contributes to TPCL slipping, $F_{ad}$ provides the pinning resistance that regulates stick–slip events. $F_{\tau}$ at the droplet base is negligible ($\sim10^{-12}~N$) relative to $F_{g,\parallel},~F_{ad}$ and can thus be ignored in our experiments\cite{al2018water}. From Eq. \ref{eqn:force balance}, we can infer that the interplay of forces like $F_{g,\parallel}$ and $F_{ad}$ shall allow a parametric control on varying the dimensional features of the patterns.

\subsection{Effect of gravity-induced asymmetry on the stripe pattern morphology}

\begin{figure}[!h]
\centering
\includegraphics[width = {\textwidth}]{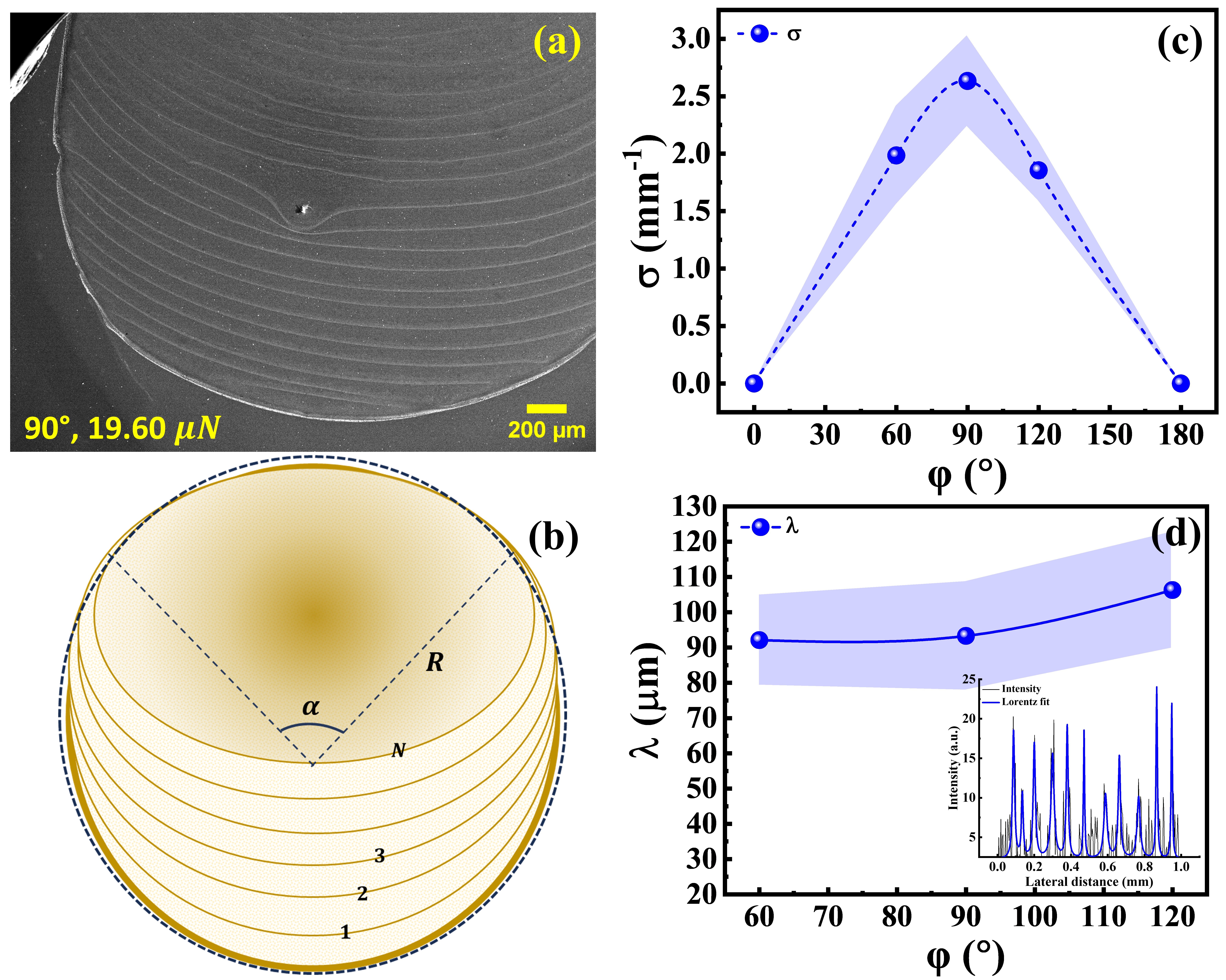}
\caption{FEG-SEM micrograph of dried AuNP droplet at $\phi=90^\circ$ (a) is shown here. Schematic (c) shows the fitting technique to obtain $\sigma,~\lambda$. Variation of stripe density $\sigma$ and stripe spacing ($\lambda$) with $\phi$ are plotted in (b,d), respectively. A representative line scan profile from the SEM image is shown in inset (d), where the peak-to-peak distances from the fitted plot are taken as the estimation of $\lambda$.}
\label{fig-sem-angle-var}
\end{figure}

The gravitational drag force $F_{g,\parallel}$ acting on a tilted drop is directly proportional to the substrate inclination. To explore how changes in $F_{g,\parallel}$ influence the stripe dimensions and achieve tunability, we performed droplet evaporation experiments at different $\phi$ values ranging from $0^\circ-180^\circ$.  We observe the formation of stripe patterns for all titled substrates ($\phi=60$\textdegree, $90$\textdegree, $120$\textdegree), while coffee-ring and coffee-eye structures formed at  $\phi=0$\textdegree and $\phi=180$\textdegree, respectively [Figure \ref{fig:all-data}(a-e)]. Figure \ref{fig-sem-angle-var}(a) shows the stripe patterns corresponding to $\phi=90^\circ$ where maximum influence of gravity is observed ($F_{g,\parallel}=F_{g,max}~\text{at}~\phi=90^\circ$). 
To quantify the dimensional features of the stripe patterns, we introduce two parameters: the number of stripes per unit length or the linear stripe density ($\sigma$), and the inter-stripe spacing ($\lambda$). As it is observed that the centers of the striped deposits can be different (slips inwards) when we count the stripes away from the TPCL, we developed a fitting process to estimate $\sigma$, normalized with the corresponding deposit perimeters, as shown in Figure \ref{fig-sem-angle-var}(b). From the circular fitting of the deposit pattern image, the radius $R$ is obtained. Two radii are drawn from the intersection of the last ($N^{th}$) stripe and the circumference to form the central angle $\alpha$. For a total of $N$ stripes, the effective length becomes $=\frac{2\pi-\alpha}{2\pi}\times2\pi R=R(2\pi - \alpha)$, where $\alpha$ is in radians. The linear stripe density $\sigma$ is therefore given by, 
\begin{equation}
\label{eq:calc-lambda}
    \sigma=\frac{N}{R(2\pi - \alpha)}
\end{equation}
Stripe spacing is calculated as the centre-to-centre separations ($\lambda_i$) between the stripes along the radial direction [Figure \ref{fig-sem-angle-var}(d) inset], and given by $\lambda=\frac{1}{N}\sum_{i=1}^N \lambda_i$. Figure \ref{fig-sem-angle-var}(c) evidently depicts the non-monotonic  variation in $\sigma$ with $\phi$. The trend in $\sigma$ is approximately sinusoidal, at par with the experimental inclinations ($\sigma\sim \sin\phi$). The vertically aligned droplet, which experiences the maximum gravitational drag ($F_{g,max}=19.6~\mu N$), produces the maximum linear stripe density. The stripe spacing $\lambda$ exhibits a weak dependence on $\phi$, with average estimate of $97.5\pm14.6~\mu m$ [Figure \ref{fig-sem-angle-var}(d)]. Although the overall trend suggests a slight increase in spacing with increasing $\phi$, the large error bars indicate a weak dependence on $\phi$, in turn, $F_{g,\parallel}$. The dependence of gravitational force on inclination angle allows for tunability in stripe density, yet the results show that the stripes remain nearly equally spaced. This non-trivial observation prompted us to examine the asymmetric droplet shapes more closely for better insight.  

\begin{figure*}[h]
\centering
\includegraphics[width = {0.8\textwidth}]{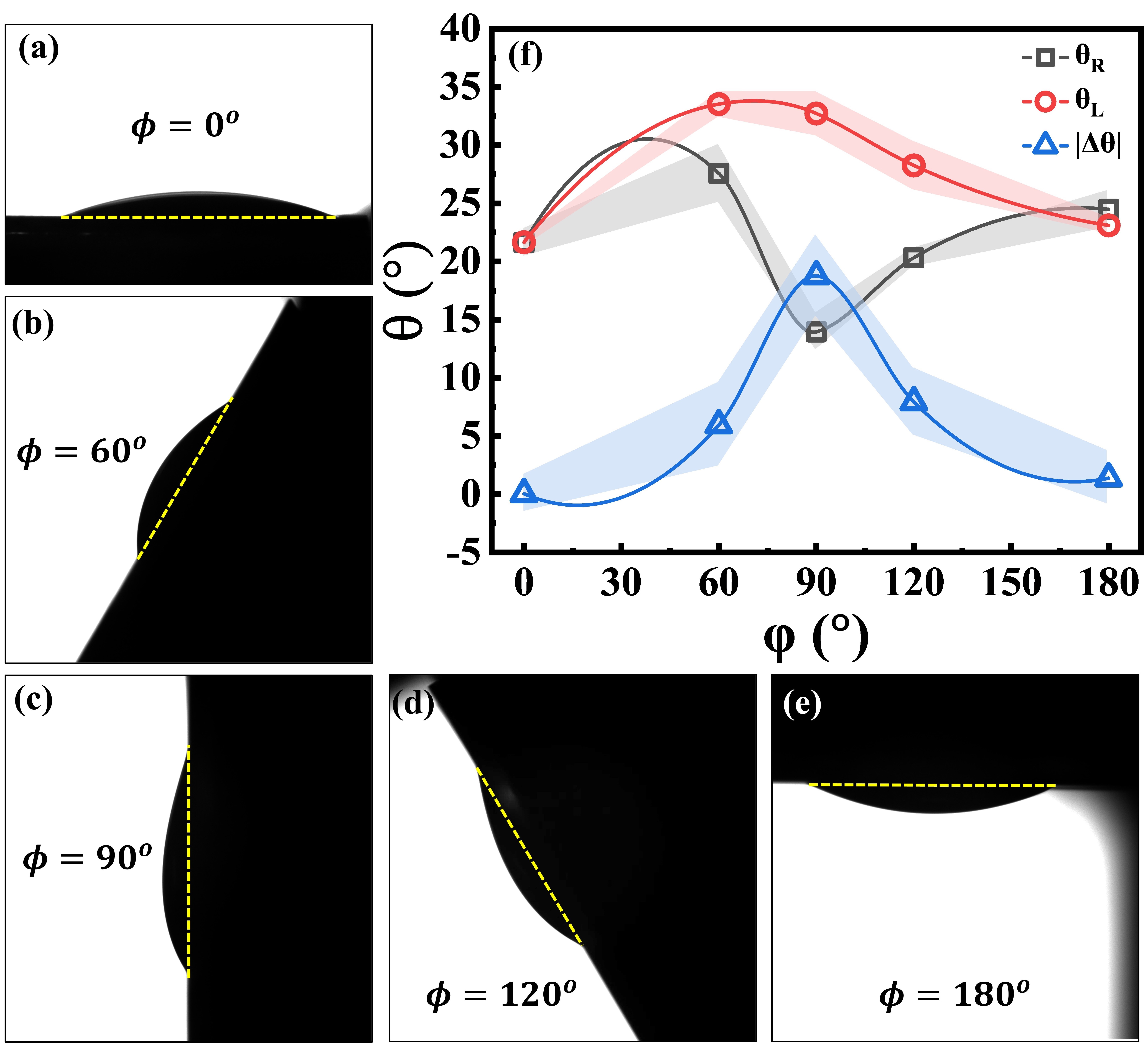}
\caption{Drop shape images recorded by optical tensiometer with varying gravity effect by changing the inclination angle $\phi$ (a-e). Dependence of $\theta_L,\theta_R$ and $\Delta\theta$ on $\phi$ are shown in (f).}
\label{fig-anglevar}
\end{figure*}
We analyzed the droplet shapes (\textit{in situ}) to understand the effect of gravitational forces on the droplet instability and hence the AuNP stripes. Figure \ref{fig-anglevar}(a-e) shows drop shape images at various tilting angles, $\phi \in [0^\circ,180^\circ]$ for AuNP droplets of concentration $C_{np}=15~nM$. These droplet images were taken using a camera connected to the tensiometer, immediately after the substrate inclination was introduced. Asymmetric drop shape analysis provides information about derived parameters like  $\theta_L$ and $\theta_R$, as well as intrinsic parameters like droplet radius ($R$)\cite{dunlop2020identities}. As expected, the droplet's TPCL for the sessile ($\phi=0$\textdegree) and pendant drop ($\phi=180$\textdegree) configuration remains pinned at both ends with $\theta_L\approx\theta_R$ \cite{srivastava2020dual,mondal2018patterns}. The other configurations result in asymmetric droplet shapes as evident from the estimates of $\theta_L$, $\theta_R$, and non-zero $\Delta\theta$ shown in Figure \ref{fig-anglevar}(f). While the sessile and pendant drop configurations do not have much hysteresis ($\Delta\theta\approx0^\circ$), the effect of gravity in the other configurations is prominent. As $\phi$ increases from $0^\circ$ to $90^\circ$, $\theta_R$ decreases, and more liquid volume is pushed towards the inclined end, creating a bulge-like profile [Figure \ref{fig-anglevar}(b-d)]. The sinusoidal trend of the direct gravity effect ($\Delta\theta\sim\sin\phi$) is evident from the similar hysteresis for supplementary angles like $\phi=0$\textdegree, 180\textdegree~ and $\phi=60$\textdegree, 120\textdegree~[Figure \ref{fig-anglevar}(f)]. The largest hysteresis is obtained at $\phi=90$\textdegree, suggesting the maximum downward drag ($F_{g,max}=19.6~\mu N$) due to the gravitational component. By correlating $\sigma$ with $\Delta \theta$ [refer to Figure \ref{fig-sem-angle-var}(c), \ref{fig-anglevar}(f)], we can quantitatively conclude that the larger the hysteresis due to gravitational drag, the higher will be the stripe density. The effect of gravitational drag is also evident from the estimation of coffee-ring widths (CRW) as presented in Figure \ref{CRW-gravity}. The CRW at the left, pinned TPCL ($w_L$), is always estimated to be equal to or higher than the rear/receding side width ($w_R$). The similar trends observed in CRW and $\Delta\theta$ with $\phi$, confirm that gravity primarily governs the formation of these dried patterns. We demonstrate that while tunability in stripe density is achievable, the gaps between stripes remain largely unaffected by gravity’s influence. 

\subsection{Effect of interfacial tension on stripe pattern formation}

\begin{figure*}[!h]
\centering
\includegraphics[width = {\textwidth}]{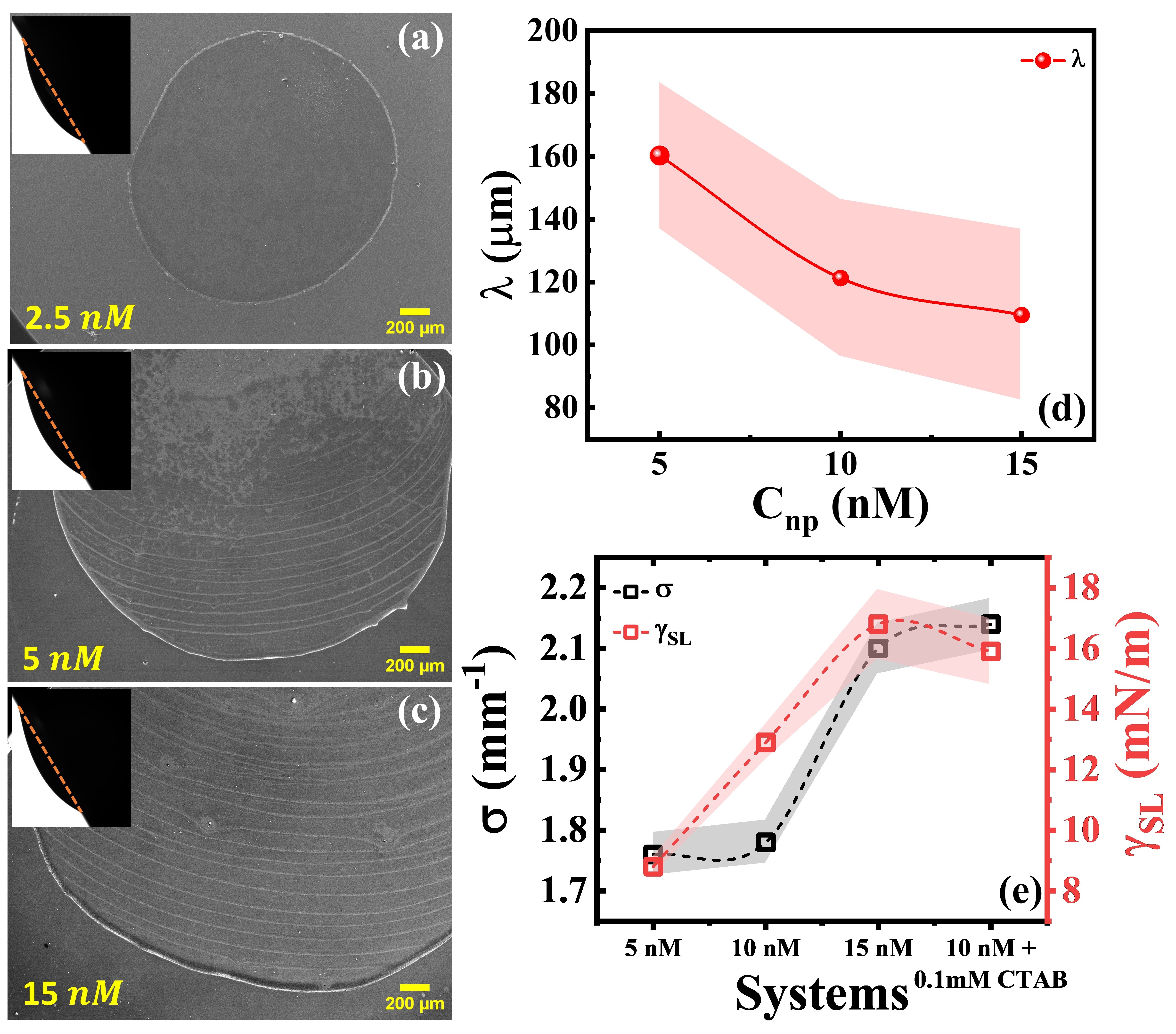}
\caption{\textit{Ex situ} dried patterns of $2.5,5,15~nM$ AuNP droplets at $\phi=120^\circ$ are shown in (a-c) [insets showing \textit{in situ} droplet profiles at $t=0~s$]. Stripes start to form only for $C_{np}\geq 5~nM$. Stripe spacing $\sigma$ decreases with increasing $C_{np}$ (d), indicating enhanced pinning frequency. Consequently, with an increase in $C_{np}$, stripe density $\lambda$ increases, and the direct effect of $\lambda$ and $\gamma_{SL}$ is established for all the systems (e).}
\label{fig-conc-var}
\end{figure*}

As the stick-slip events of TPCL are a mechanical as well as interfacial phenomenon, we aimed at manipulating the TPCL dynamics (pinning-depinning events) to introduce tunability of the deposited stripes in terms of $\lambda$. As the pinning force on the TPCL is dependent on the interfacial tension, varying the particle concentration is the simplest way to achieve that tunability. The AuNP concentration $C_{np}$ is directly related to $\gamma_{LV}$, which in turn relates to $\gamma_{SL}$ [Eq. \ref{eqn:Young}].  We studied the concentration effect on the stripe pattern at a fixed inclination ($\phi=120^\circ$). Figure \ref{fig-conc-var}(a-c) shows the dried deposits for $C =2.5~nM$ (no stripe), and $5~nM,15~nM$ (with stripes). The corresponding FEG-SEM images for AuNP droplets with $C_{np}\in[1~nM,~15~nM]$ dried in the same tilted configuration are shown in Figure \ref{fig:all-data}(f-j). Our findings indicate that stripe formation only occurs at concentrations of $5~nM$ or higher; concentrations below this threshold do not result in stripes. For the estimates of $\sigma$ and $\lambda$ we only considered $C_{np}\geq 5~nM$. Interestingly, when the stripes were produced through the manipulation of surface tension via the change in nanoparticle concentration, we observed a clear inverse relationship between $\lambda$ and $C_{np}$ [Figure \ref{fig-conc-var}(d)], as opposed to the systems where patterns were produced only via a change in $\phi$. As $C_{np}$ increases from $5 ~nM$ to $15 ~nM$, the average stripe spacing decreases from $\sim160~\mu m$ to $\sim100~\mu m$. Concurrently, $\sigma$ shows a steady increase with increasing $C_{np}$ [Figure \ref{fig-conc-var}(e)], as can be expected from the closely-spaced stripes. We found that increasing the particle concentration allowed clear controllability in $\sigma$ and $\lambda$. The gravitational drag, $F_{g,\parallel}$ on these droplets was kept constant, which reflects in the negligible change in $\Delta \theta$. This is also evident from the slight change in Bond number ($0.30-0.36$), which is the ratio of gravity to surface tension force (refer to SI, Figure \ref{fig:bond-number}(a,b)). Furthermore, for concentrations below the threshold of $C_{np}<5~nM$, we hypothesize that after the initial pinning, the colloidal droplet lacks the minimum number of particles necessary to sustain further TPCL pinning; consequently, all particles are deposited within the coffee ring itself. Above the critical concentration of $5~nM$, there are sufficient particles in the system to enable multiple stick-slip motions of the TPCL following the initial coffee ring formation. This observation is in line with the work of Srivastava \textit{et al.} where stripe patterns were obtained even at $C_{np}=2.5~nM$ for a sessile drop\cite{srivastava2020dual}. In our experiments, a higher threshold concentration ($C_{np} \geq 5~nM$) is required for stripe formation on inclined substrates. This suggests that the gravitational drag along the inclined substrate suppresses repeated stick--slip events at lower nanoparticle concentrations, where TPCL pinning is insufficient.
\par 
\par

To understand the origin of the observed surface morphology of the deposit pattern shown in Figure \ref{fig-sem-angle-var}(a) and \ref{fig-conc-var}(a-c) obtained by varying $\phi$ and $C_{np}$ respectively, let us have a look at the adhesion force term ($F_{ad}$), defined in Eq. \ref{eqn:force balance}. The solid-liquid interfacial tension ($\gamma_{SL}$) is directly related to the adhesion force\cite{al2020adhesion} as,
\begin{equation}
\label{f-ad}
    F_{ad}\propto\gamma_{SL}(cos\theta_R-cos\theta_L)
\end{equation}
$\gamma_{LV}$ values for the different systems are estimated by the Pendant drop method using an optical tensiometer (refer to SI section \ref{sec:pendant-drop}), and consequently $\gamma_{SL}$ is calculated from Eq. \ref{eqn:Young}, employing $\theta_R$ for the receding end of the droplet. Interfacial tension $\gamma_{LV}$ decreases with an increase in $C_{np}$, which in turn increases $\gamma_{SL}$ [table \ref{table Fad}]. Adhesion force ($F_{ad}$), which is directly proportional to $\gamma_{SL}$, results in TPCL pinning and promotes the formation of denser stripes with increasing $C_{np}$. With only $F_{g,\parallel}$ variation, nanoparticle density was fixed, resulting in an invariant effect on the TPCL pinning, vis-\`a-vis $\lambda$. But the asymmetry in the droplet shape ($\Delta \theta$) contributes to the normalized $\sigma$, showing a sinusoidal trend. These results validate the significance of interfacial tensions in the formation of ordered stripes, while gravity remains uniform. 
\begin{table*} [!h]
\centering
\begin{tabular}{ |c|c|c|c|c| } 
 \hline
 $C_{np}~(nM)$ & $\theta_R$(\textdegree) & $\gamma_{LV}~(mN/m)$ & $\gamma_{SL}~(mN/m)$ \\ 
 \hline
5	& 23.2	& 72.7	& 8.2 \\  
 \hline
 10 & 22.1 &	67.4 & 12.6 \\ 
 \hline
 15 &	20.3 &	61.1 &	17.7 \\
 \hline
 10 + 0.1 mM CTAB	& 21.1	& 60.5	& 18.5\\
 \hline
\end{tabular}
\caption{Estimation of $\gamma_{LV}$ and $\gamma_{SL}$ with varying particle concentrations ($C_{np}$) are given below.}
\label{table Fad}
\end{table*}
Our results indicate that the dimensions of the stripe patterns, specifically $\sigma$ and $\lambda$, can be controlled by adjusting a single parameter - the surface tension of the sessile drop. Since surface tension can be modulated by altering either the particle density or surfactant concentration within the droplet, we conducted an additional experiment where the liquid–vapor surface tension ($\gamma_{LV}$) was reduced by introducing $0.1~mM$ CTAB. This concentration was carefully chosen to yield an estimate of $\gamma_{SL}$ comparable to that of the nanoparticle system with $C_{np} = 15~nM$ (without added surfactant), enabling direct comparison. As shown in Figure \ref{fig-conc-var}(e), the $\gamma_{SL}$ values in the two cases are nearly identical within the error margin. Notably, the corresponding $\sigma$ values also fall within a similar range, providing direct evidence for the role of interfacial tension in governing stripe pattern formation with controlled dimensions. Thus, we demonstrate that by fine-tuning surface tension, equivalent striped morphologies can be reproducibly achieved with sub-millimeter precision.

\section{Conclusion}
In summary, our work offers the first experimental evidence for the creation of semi-circular stripe patterns with controllable dimensions, through solvent evaporation of colloidal droplets on inclined substrates. We provide a correlation between $\sigma$, $\lambda$, and the experimental parameters such as particle concentration and interfacial tension, thus allowing us to pattern a surface with a stripe-like morphology of desired dimensions. This method shall be useful for various applications that require length scale dependence responses, such as sensing. \par
While previous studies have observed stick-slip dynamics of the TPCL on inclined substrates\cite{varagnolo2013stick,winhard2021direct}, the formation of well-defined, tunable periodic patterns has not been reported. We show that gravity-driven instability is the driving force for the stick-slip events leading to the formation of stripe patterns. The stripe density $\sigma$, mirrors the sinusoidal dependence of the gravitational drag force component on the inclination angle ($\phi$). Thus, maximum stripe density is observed at $\phi=90^\circ$, also correlating with the largest droplet shape asymmetry through $\Delta\theta$. Although gravity controls the stripe density, we have demonstrated that stripe spacing can be controlled by altering interfacial forces.  This is achieved by varying the surface tension through changes in particle concentration or by the surfactant concentration, which impact the solid-liquid interfacial tension $\gamma_{SL}$. We find that a critical nanoparticle concentration is required for stripe formation, consistent with the analytical relation developed earlier for the occurrence of stick-slip motion on evaporating sessile drop \cite{srivastava2020dual}. By revealing the intricate balance between gravity and interfacial forces in stripe formation, this work establishes a framework for programmable and scalable nanoscale patterning strategies in materials science. 

\section{Experimental Methods}
\label{sec:expt}
\subsection*{Materials}
Gold (III) chloride trihydrate - $HAuCl_{4}.3H_{2}O$ ($\geq 99.9\%$) and Trisodium citrate dihydrate - $C_{6}H_{5}Na_{3}O_{7}.2H_{2}O$ ($\geq99.0\%$) are obtained from Sigma Aldrich. Hexadecyltrimethylammonium bromide/CTAB - $C_{19}H_{42}BrN$ ($>98\%$) and Sodium borohydride - $NaBH_{4}$ ($>96\%$) were purchased from Spectrochem Pvt Ltd (India). Nitric acid ($HNO_{3}$) and hydrochloric acid ($HCl$) of EMAPRTA grade (Merck), were used for the preparation of aqua regia to clean glassware. Sulphuric acid - $H_2SO_4$ (98\%), Hydrogen peroxide - $H_2O_2$ (70\%) and Acetone (EMPARTA) were obtained from Merck. Deionized (DI) water (resistivity $18.2~M\Omega cm$) is used for all experiments.
Silicon wafers (thickness $275\pm25~\mu m$) are obtained from CEN, IIT Bombay.

\subsection*{Synthesis of AuNPs}
The CTAB-coated spherical gold nanoparticles are synthesized using the improvised protocol developed earlier based on seed-mediated growth process \cite{kumar2024shape}. Citrate-capped seeds are first synthesized, which subsequently generate CTAB-capped larger nanoparticles through ligand exchange. For the seed, $300~\mu L$ of freshly prepared ice-cold $NaBH_4$ ($0.1~M$) is introduced into a $10~mL$ solution containing $0.25~mM$ $HAuCl_4$ and sodium citrate, and the mixture is incubated for 2 hours at room temperature. In parallel, a growth solution consisting of $80~mM$ CTAB and $0.25~mM$ gold is prepared, where the temperature may be raised to $50^\circ C$ to facilitate CTAB dissolution but is subsequently cooled to room temperature prior to use. To $45~mL$ of this growth solution, $0.25~mL$ of freshly prepared $100~mM$ ascorbic acid is added under vigorous stirring, followed by the addition of $5~mL$ of the seed solution. Stirring is done for 10 minutes until the solution turns dark red, after which it is incubated overnight at room temperature. The final growth solution is centrifuged twice at 14000 rpm for 15 minutes to eliminate excess CTAB and other contaminants. Characterizations of the synthesized particles are provided in SI section \ref{sec:particle charact}.

\subsection*{Evaporative assembly setup and analysis of deposition patterns}
For droplet deposition, a single-side polished 2" \textit{Si} wafer was cut into $1~cm\times1~cm$ pieces and cleaned using acetone, Piranha solution (1:1), and a stream of D.I water and dried under $N_2$ gas. For \textit{in situ} evaporation profile measurements, micro-droplets of nanoparticle suspensions ($2~\mu L$) were dropcast on freshly cleaned substrate and images of the droplets were recorded through a camera connected to an optical tensiometer (Attension Theta Flex, Biolin Scientific). The evaporation of an inclined colloidal droplet is recorded using a tensiometer stage that can be rotated from $0^\circ-360^\circ$. For this experiment, the inclination/tilting angle ($\phi$) is kept between $0^\circ-180^\circ$ due to the oscillating nature of sinusoidal functions. Each set of evaporation experiment is repeated several times ($\geq4$) to ensure reproducibility and the standard deviations in the measurements are presented as error bars in corresponding plots. \textit{In situ} measurements like contact angle ($\theta$), drop diameter ($D$), and other derived parameters are analyzed by the OneAttension software (Biolin Scientific) using Young-Laplace equation fitting. After the droplet is fully dried, \textit{ex situ} measurements such as FEG-SEM imaging (JEOL JSM-7600F) are done for surface morphology investigation. ImageJ software is used for analyzing SEM micrographs\cite{schneider2012nih}.

\section*{Supporting Information}
Supporting information includes nanoparticle characterizations, complete inclined-droplet \textit{ex situ} datasets, CRW analysis, $\gamma_{LV}$ estimation, contact angle and Bond number variations with $C_{np}$.

\begin{acknowledgement}
The authors thank Prof. A. Sain (Dept. of Physics, IIT Bombay) for helpful discussions. The authors acknowledge CSIF IIT Bombay for FEG-SEM and FEG-TEM(200kV/300kV) facilities. 
SB acknowledges the financial support from CRNTS and IRCC, IIT Bombay.
SK acknowledges financial support from the CSIR, India, and IRCC-IIT Bombay. SS acknowledges support from UGC-DAE CSR (CRS/2022-23/03/876), India and IIT Bombay.

\end{acknowledgement}
\section*{Conflict of interest}
The authors declare no conflict of interest.
\section*{Data Availability Statement}
The data that support the ﬁndings of this study are available from the corresponding author upon reasonable request.
\section*{Keywords}Inclined droplet, evaporative assembly, gold nanoparticle, periodic stripes, tunable assembly. 

\bibliography{ref}

\end{document}


\pagebreak
\section{Characterizations of AuNPs}
\label{sec:particle charact}
\begin{figure*} [h]
\centering
\includegraphics[width = {0.75\textwidth}]{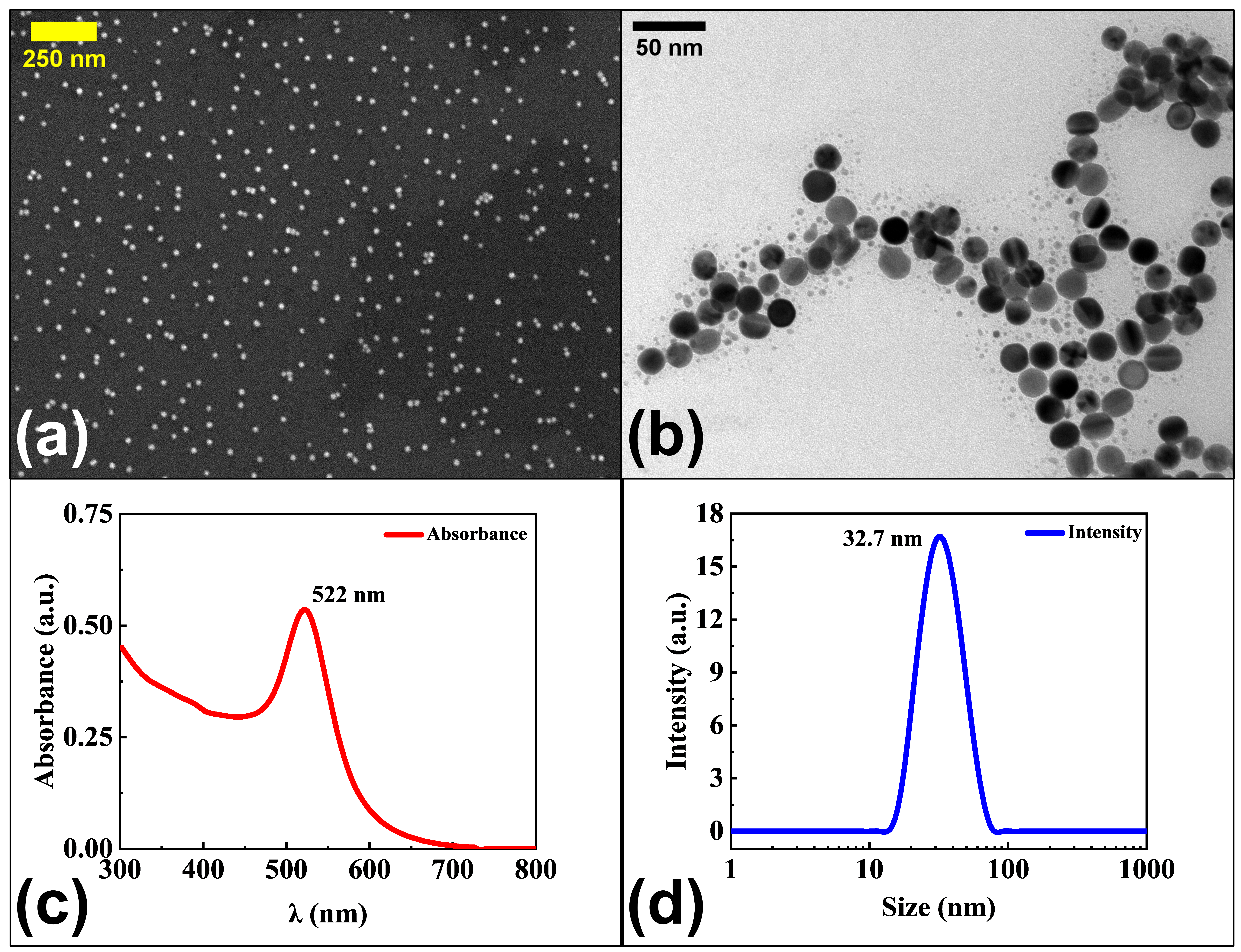}
\caption{High magnification SEM (a) and TEM image (b) showing spherical CTAB-AuNPs of size $\sim 15~nm$. UV-Vis absorption spectra (c) show a maximum absorbance peak at $522~nm$. DLS shows the hydrodynamic size of spherical AuNPs $\sim 33~nm$ (d).}
\label{ctab-aunp-char}
\end{figure*}

The synthesized AuNPs are characterized via FEG-SEM and FEG-TEM imaging, UV-Visible spectroscopy, and DLS. Figure \ref{ctab-aunp-char}(a,b) shows the SEM and TEM micrographs for monodisperse AuNPs of size $\sim15~nm$, respectively. Figure \ref{ctab-aunp-char}(c) shows the UV-Vis spectra of AuNP colloidal solution, exhibiting an SPR peak at $522~nm$. DLS data of the colloidal particles shows a hydrodynamic size of $\sim33~nm$, confirming the existence of a hydrated CTAB bilayer around the cores of AuNPs [Figure \ref{ctab-aunp-char}(d)].

\section{Dried patterns from inclined droplets: All experimental data}
\begin{figure} [h]
\centering
\includegraphics[width = {\textwidth}]{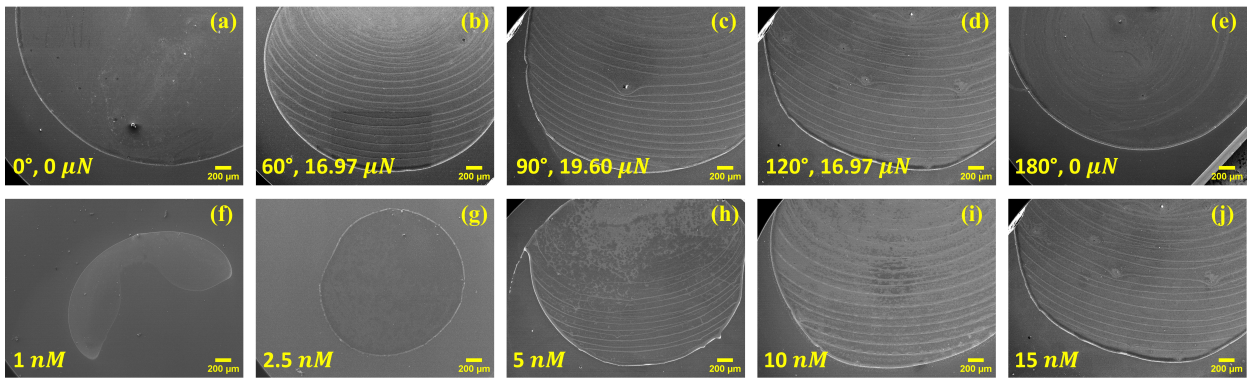}
\caption{Top panel (a-e) shows the FEG-SEM images of dried deposits of AuNP droplet ($C_{np}=15~nM$) for $\phi=0^\circ,60^\circ,90^\circ,120^\circ, \text{and }180^\circ$. Bottom panel (f-j) shows the concentration dependence on dried deposits of AuNP droplets at $\phi=120^\circ$.}
\label{fig:all-data}
\end{figure}

Figure \ref{fig:all-data}(a-e) shows the dependence of gravity on the formation of stripe patterns through the evaporative self-assembly of AuNP droplets. The concentration of AuNP is kept fixed at $15~nM$ in these experiments. We observe that coffee-ring [Figure \ref{fig:all-data}(a)] and coffee-eye [Figure \ref{fig:all-data}(e)] structure is formed for the sessile ($\phi=0^\circ$) and pendant drop ($\phi=180^\circ$), respectively. All the other inclined drops formed striped patterns after drying [Figure \ref{fig:all-data}(b-d)].
The effect of particle concentration, as well as the interfacial tension, is explored to understand the stripe pattern formation. A critical concentration $C_{np}=5~nM$ is identified, beyond which the stripes start to form. As can be seen in Figure \ref{fig:all-data}(f-j), coffee-rings are formed without any stripes for particle concentration $C_{np}<5~nM$. For $C_{np}\geq 5~nM$, stripes are formed with increasing stripe density [refer to Figure \textcolor{blue}{4}(e) in the main text].

\section{Gravitational effect on the coffee-ring widths (\texorpdfstring{$w$}{w})}

\begin{figure} [hbpt]
\centering
\includegraphics[width = {0.6\textwidth}]{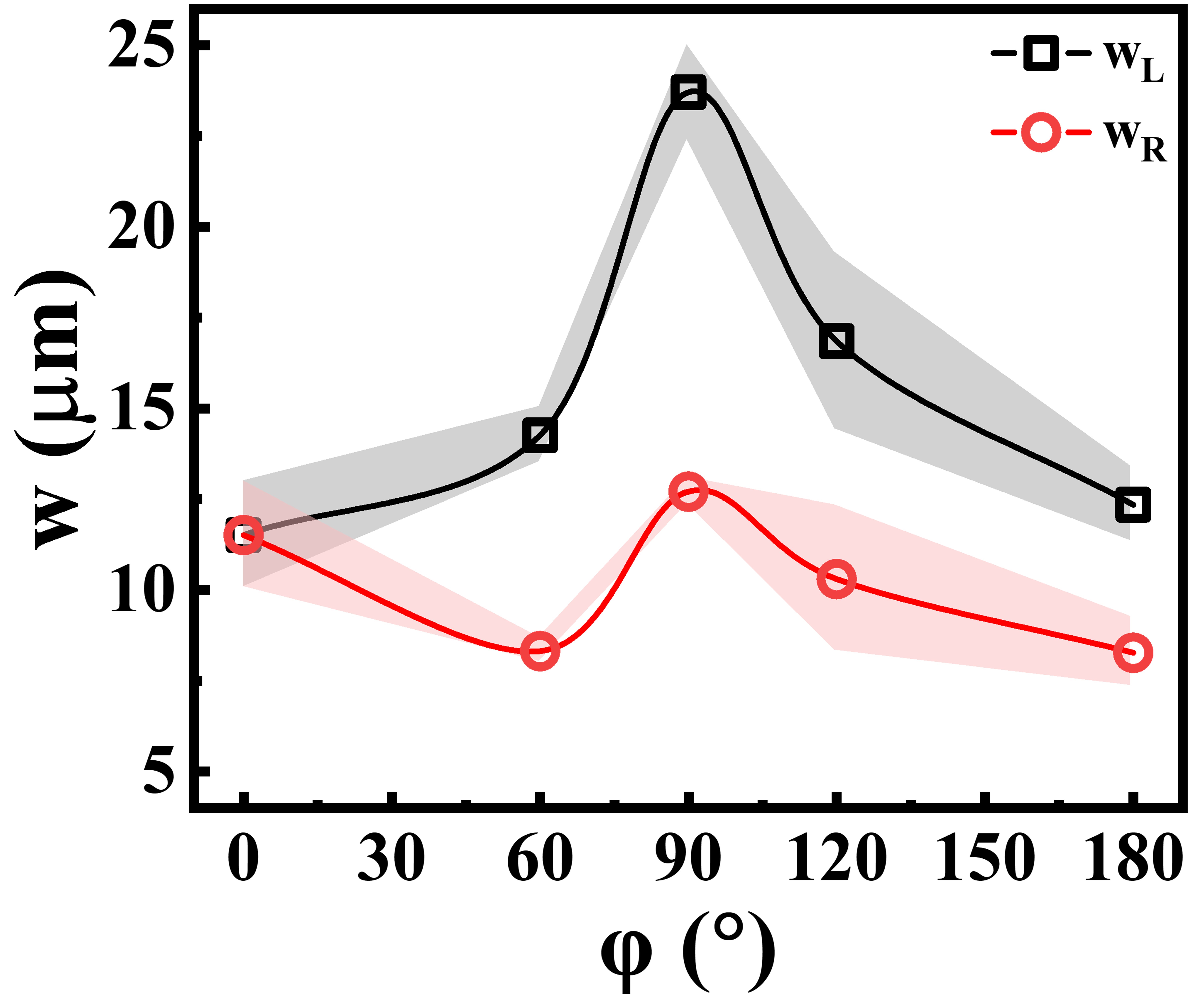}
\caption{Coffee-ring widths on the left/pinned ($w_L$) and right/receding side ($w_R$), are plotted as a function of $\phi$. }
\label{CRW-gravity}
\end{figure}
The effect of gravitational drag is evident in the estimation of coffee-ring widths ($w_L,w_R$) as presented in Figure \ref{CRW-gravity}. The width of the coffee-ring at the left, pinned TPCL ($w_L$), is always estimated to be equal to or higher than the right/receding side width ($w_R$). The trend in CRW is similar to $\sin\phi$ or $\Delta\theta$, which directly establishes the effect of gravity in the formation of the dried patterns.

\section{Estimation of liquid-vapor interfacial tension (\texorpdfstring{$\gamma_{LV}$}{gammaLV}) for different systems}
\label{sec:pendant-drop}

\begin{figure} [h]
\centering
\includegraphics[width = {0.3\textwidth}]{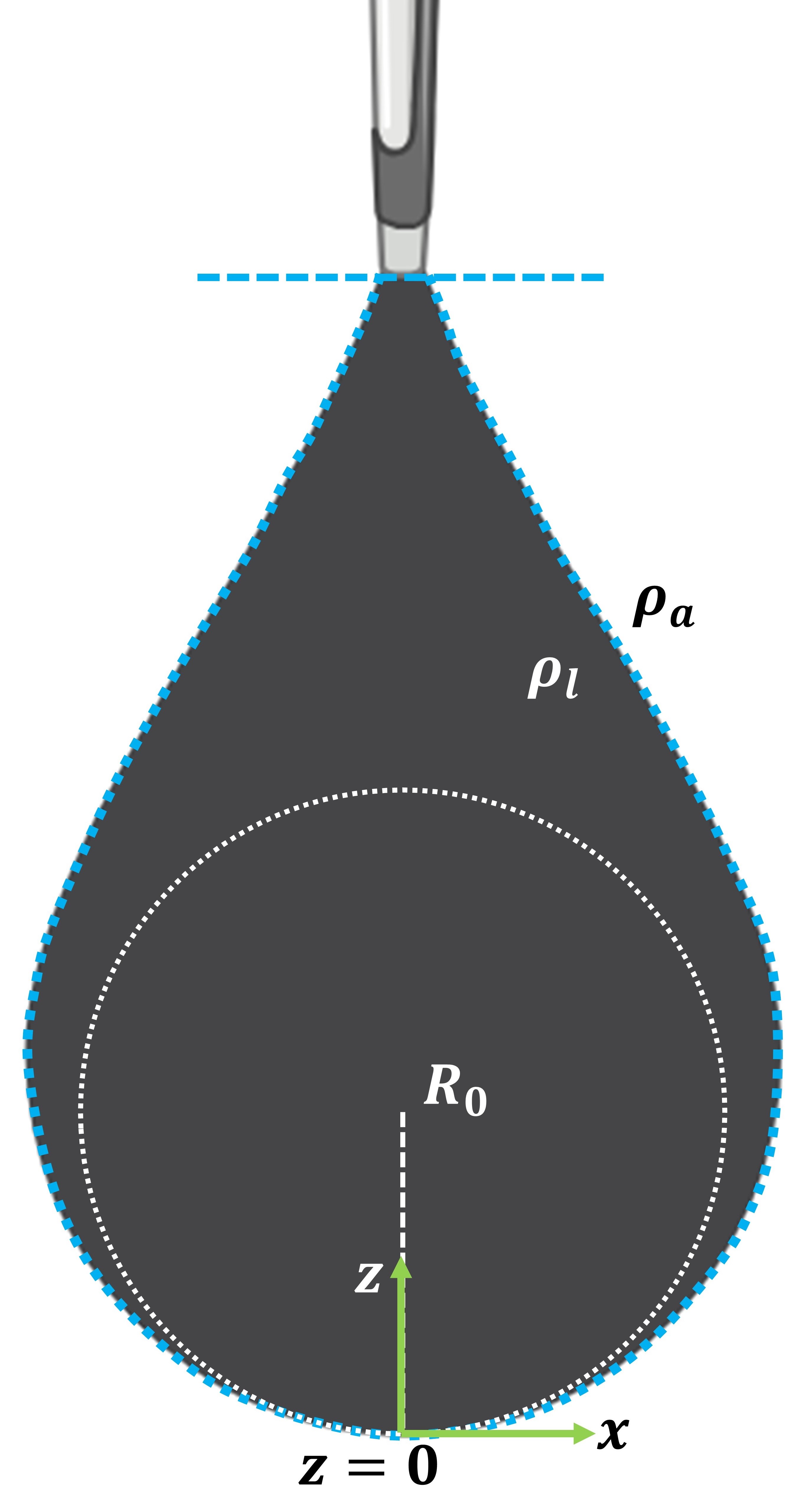}
\caption{Schematic of a pendant drop with Young-Laplace fitting for the estimation of $\gamma_{LV}$.}
\label{fig:pendant}
\end{figure}

Liquid–vapor interfacial tension ($\gamma_{LV}$) is measured using the pendant drop method, which is widely applied to colloidal suspensions at surfaces and interfaces \cite{pan2024interfacial}. 
\begin{equation}
    \gamma\Big(\frac{1}{R_1}+\frac{1}{R_2}\Big)=\Delta p_0 - \Delta \rho gz 
\label{eq:YL}
\end{equation}
For a pendant drop at equilibrium [Eq. \ref{eq:YL}], the Laplace pressure across the interface is given by $\Delta p \equiv \Delta p_0 - \Delta\rho gz$. Here, $\Delta p$ denotes the pressure difference at $z=0$ relative to the hydrostatic pressure ($\rho g z$). $\Delta\rho$ is the density difference between the liquid ($\rho_l$) and the surrounding phase ($\rho_a$), while $g$ is the gravitational acceleration. The surface (or interfacial) tension is experimentally determined by fitting the drop profile geometrically [Figure \ref{fig:pendant}] using Eq. \ref{eq:YL}.

\section{Variation in contact angles (\texorpdfstring{$\theta$}{theta}) and Bond number (\texorpdfstring{$B_o$}{Bo}) with AuNP concentrations (\texorpdfstring{$C_{np}$}{Cnp})}

\begin{figure} [h]
\centering
\includegraphics[width = {0.911\textwidth}]{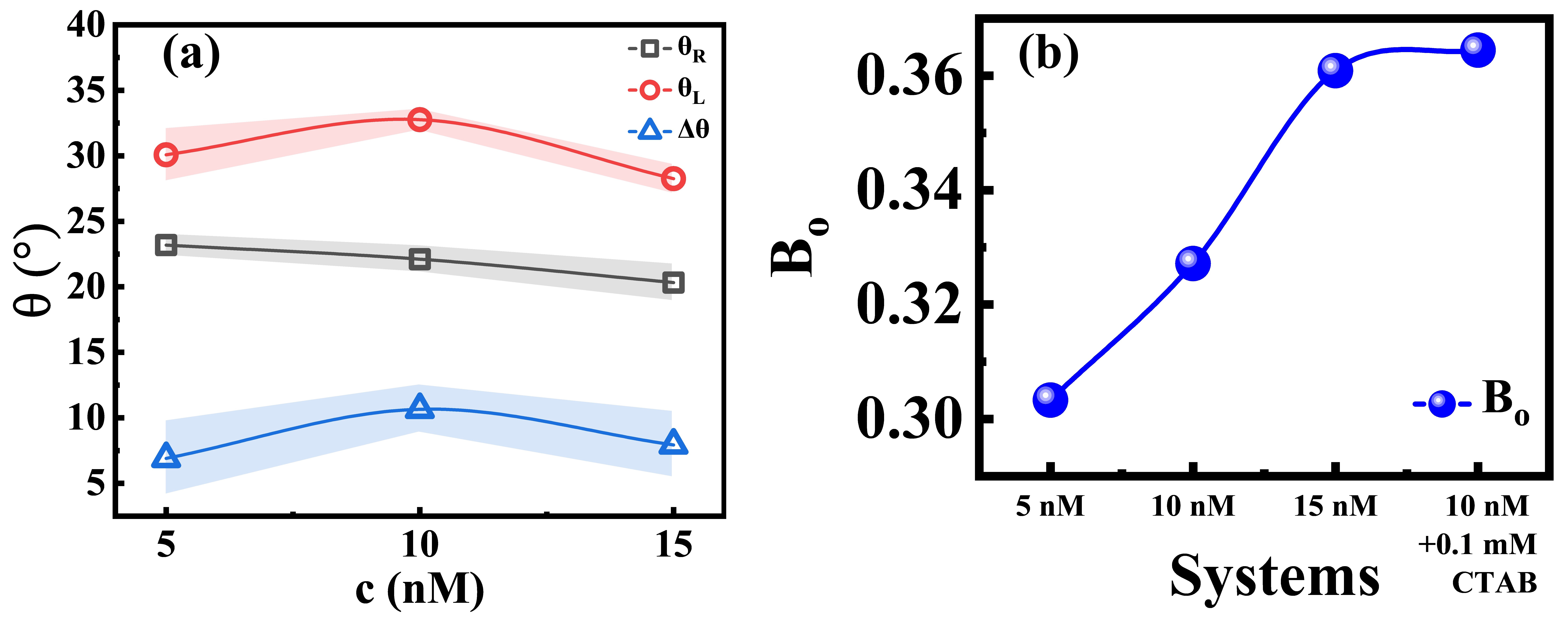}
\caption{Dependence of $\theta_L,\theta_R$ and $\Delta\theta$ with AuNP concentration are shown in (a). Calculated $B_o$ for different systems are plotted in (b).}
\label{fig:bond-number}
\end{figure}

As observed in our experiments, stripe patterns are only formed beyond a critical particle concentration of $5~nM$. Figure \ref{fig:bond-number}(a) shows the variations in $\theta_L,\theta_R$, and $\Delta\theta$ with increasing $C_{np}$. As the droplets are pinned from the beginning, $R_0$ is approximately the same for all cases at constant gravitational drag ($\phi=120^\circ, ~F_{g,\parallel}=16.97~\mu N$). $\theta_R$ values are always $<25^\circ$, from where the stick-slip motion is happening. A constant trend in the $\Delta\theta$ values suggests equivalent gravitational effect on all the drops. The decline in the $\theta_R$ denotes the influence of particle concentration ($C_{np}$), which can be correlated to the decrease in $\gamma_{LV}$.
\par
Bond number ($B_o$) is the ratio of gravitational force to the surface tension force, and is defined as $B_o=R^2\rho g/\gamma_{LV}$. Figure \ref{fig:bond-number}(b) shows the variation in $B_o$ from for different systems, signifying almost equivalent effect of gravity ($B_o\approx0.30-0.36$) while surface tension dominates.

\bibliography{ref}